\documentclass[letterpaper]{article} 
\usepackage[T1]{fontenc}
\usepackage{aaai25}  
\usepackage{times}  
\usepackage{helvet}  
\usepackage{courier}  
\usepackage[hyphens]{url}  
\usepackage{graphicx} 
\def\UrlFont{\rm}  
\usepackage{natbib}   
\usepackage{caption}  
\usepackage{dirtytalk}

\usepackage{algorithm}
\usepackage{algorithmic}
\usepackage{newfloat}
\usepackage{listings}
\DeclareCaptionStyle{ruled}{labelfont=normalfont,labelsep=colon,strut=off} 
\floatstyle{ruled}
\newfloat{listing}{tb}{lst}{}
\floatname{listing}{Listing}
\newcommand{\sayone}[1]{`#1'}
\title{Bias is a Math Problem, AI Bias is a Technical Problem: 10-year Literature Review of AI/LLM Bias Research Reveals Narrow [Gender-Centric] \\ Conceptions of \say{Bias}, and Academia-Industry Gap }

\author{
    Sourojit Ghosh,
    Kyra Wilson
}
\affiliations {
    University of Washington, Seattle\\
    \{ghosh100, kywi\}@uw.edu
}

\begin{document}

\maketitle

\begin{abstract}

The rapid development of AI tools and implementation of LLMs within downstream tasks has been paralleled by a surge in research exploring how the outputs of such AI/LLM systems embed biases, a research topic which was already being extensively explored before the era of ChatGPT. Given the high volume of research around the biases within the outputs of AI systems and LLMs, it is imperative to conduct systematic literature reviews to document throughlines within such research. In this paper, we conduct such a review of research covering AI/LLM bias in four premier venues/organizations -- *ACL, FAccT, NeurIPS, and AAAI -- published over the past 10 years. Through a coverage of 189 papers, we uncover patterns of bias research and along what axes of human identity they commonly focus. The first emergent pattern within the corpus was that 82\% (155/189) papers did not establish a working definition of \say{bias} for their purposes, opting instead to simply state that biases and stereotypes exist that can have harmful downstream effects while establishing only mathematical and technical definition of bias. 94 of these 155 papers have been published in the past 5 years, after \citet{blodgett2020language}'s literature review with a similar finding about NLP research and recommendation to consider how such researchers should conceptualize bias, going beyond strictly technical definitions. Furthermore, we find that a large majority of papers -- 79.9\% or 151/189 papers -- focus on gender bias (mostly, gender and occupation bias) within the outputs of AI systems and LLMs. By demonstrating a strong focus within the field on gender, race/ethnicity (30.2\%; 57/189), age (20.6\%; 39/189), religion (19.1\%; 36/189) and nationality (13.2\%; 25/189) bias, we document how researchers adopt a fairly narrow conception of AI bias by overlooking several non-Western communities in fairness research, as we advocate for a stronger coverage of such populations. Finally, we note that while our corpus contains several examples of innovative debiasing methods across the aforementioned aspects of human identity, only 10.6\% (20/189) include recommendations for how to implement their findings or contributions in real-world AI systems or design processes. This indicates a concerning academia-industry gap, especially since many of the biases that our corpus contains several successful mitigation methods that still persist within the outputs of AI systems and LLMs commonly used today. We conclude with recommendations towards future AI/LLM fairness research, with stronger focus on diverse marginalized populations.

\end{abstract}

\section{Introduction}

Though Artificial Intelligence (AI) is by no means a technology novel to the 21$^{st}$ century, the development of the AI assistant ChatGPT powered by the Large Language Model (LLM) GPT-3 and its subsequent release for public usage in November 2022 spurred an era of promise and heralded the still-ongoing AI summer. As billions of dollars are continually poured across the world into the design and adoption of novel AI systems and tools leveraging multimodal LLMs in downstream tasks alongside the promulgation of new policies and legislations such as Executive Order 14179: \textit{Removing Barriers to American Leadership in Artificial Intelligence} (\citeyear{EO14179}) in the US, there has also been a steep rise in LLM-related research throughout this period of time \cite{pang2025understanding} towards technical innovation and expanding the scope of tasks which AI systems and LLMs can accomplish. 

In addition, a significant proportion of this research has focused on the various ways in which the outputs of AI systems and LLMs exhibit bias towards traditionally marginalized populations \cite{gupta2023sociodemographic, wan2024survey}. This is a critical angle of research exploring the sociotechnicality of AI systems and LLM-powered tools, recognizing that the presence of negative stereotypes within their outputs against populations historically marginalized in society and oppressed through existing societal structures brings further harm upon them and exacerbates their marginalization. However, past research both in and adjacent to AI/LLMs, namely that of natural language processing (NLP), has demonstrated that researchers focusing on the nuances of biases within system outputs commonly display purely technical understandings of \say{bias} in their research \cite[e.g,][]{blodgett2020language}. The combination of the rapid publication of AI/LLM bias research \cite{pang2025understanding} and a previously-recognized narrow conception of \say{bias} by technical researchers in related fields necessitates an exploration of the body of work in this field.   

In this paper, we perform a systematic literature review covering work published in four premier venues/organizations --- *ACL, FAT*/FAccT, NIPS/NeurIPS, and AAAI --- exploring emergent trends within bias and fairness research in AI systems and LLMs published within the last 10 years. Our coverage includes 189 full papers, curated through computational and manual screening, as we make the following three contributions to the field:

\begin{enumerate}
    \item Firstly, we observe that 82\% (155/189) papers studying \say{bias} in AI systems or LLMs do not actually define \say{bias}, and 74 of these 136 have been published in the last 5 years during the massive surge of LLM/AI fairness research. This is particularly concerning in light of \citet{blodgett2020language}'s similar finding 5 years ago, as it is indicative that researchers in the field have not since taken their call to heart. The lack of a definition of \say{bias} in a paper studying \say{bias} poses real problems, and we provide recommendations for researchers to adopt more than  mathematical conceptions of bias in their work. 
    
    \item We document a strong focus on `gender bias' within the set of papers surveyed within our dataset, with 79.9\% (151/189) papers demonstrating such a focus. Alongside a focus on gender  (most prominently, the associations of gender with occupations, as in 19.6\% or 37/189 papers), 30.2\% (57/189) researchers also studied race/ethnicity bias (often equating the two), 13.2\% (25/189) studied nationality bias, 20.6\% (39/189) focused on age bias, and 19.1\% (36/189) researchers worked on religion bias. Our findings highlight a concerning trend within the field that research around AI bias and fairness is often restricted to axes of identity perceived globally and interpreted through Western lenses (e.g., gender and occupation bias research often focuses based on US occupations data, examples of racial bias in word embeddings refers to African-Americans as criminals, etc.). This pattern has been alluded to in the past \cite[e.g.,][]{dev2023building, ghosh2024interpretations, qadri2023ai}, but not explicitly demonstrated through such a coverage of the field as ours. We highlight potential avenues to pursue research around historically marginalized communities often overlooked in AI-fairness research, to achieve stronger coverage of users and communities producing or consuming AI outputs.  
    
    \item Finally, it is evident across the corpus of papers that though researchers excel at innovative and effective debiasing techniques across the aforementioned axes of identity and providing technical solutions towards fairer AI systems, only 10.6\% (20/189) papers contain any actionable implementation techniques of their findings within real-world AI systems. This is highlighted by an absence of potential design recommendations or mention of implementation procedures in papers, beyond making codebases and datasets publicly available for academic transparency. This finding is particularly salient because it reflects a glaring academia-industry gap: at the same time as a large volume of research is being published that is producing several methods to mitigate different biases which are demonstrated to be highly successful, real-world AI systems and LLMs commonly-used in downstream tasks contain and perpetuate such biases within their outputs. We call for stronger collaboration between academic researchers and industry developers of AI systems, and advocate for academics to include within their writing strategies for the applications of their work within commonly-used AI systems.
    
\end{enumerate}

\section{Background}

\subsection{Literature Reviews}

A literature review is a common and effective way of studying the state of a field and the various patterns of research within it, allowing for a collective understanding of salient threads of research and the identification of gaps to be filled with further research \cite{knopf2006doing}. Literature reviews are commonly of two types -- systematic literature reviews that meticulously follow replicable protocols for screening and evaluation \cite{kitchenham_procedures_2004}, and semi-systematic literature reviews that afford more flexibility and subjectivity within curation and review processes \cite{snyder2019literature}. A few recent examples include \citet{mack2021we}'s work exploring the field of Accessibility research provided a discursive overview of the field in terms of preferred communities of focus, popular research methods, the role of nondisabled participants and caregivers/specialists, and common research questions explored; \citet{birhane2022values}'s survey of the state of the art in Machine Learning research within the top 100 most `influential' papers (determined by highest citation count), finding that the research broadly values Performance, Generalization, Efficiency, Building on Past Work, and Novelty, with little to no discussion of potentially harmful outcomes or negative applications of the work; and \citet{pang2025understanding}'s exploration of LLM-related research within the CHI Conference on Human Factors, exploring the domains of HCI research where LLMs have been applied, the various roles performed by LLMs in such research and resultant types of contributions, and the limitations and risks of using LLMs as acknowledges by HCI researchers. 

Such literature reviews are also common in NLP \cite[e.g.,][]{gao2023retrieval, guo2024large, zhao2023survey}. Our interests in exploring research around \say{bias} in LLMs/AI systems is perhaps closest to \citet{bansal2022survey}'s or \citet{gupta2023sociodemographic}'s explorations of types/sources of [socio-demographic] biases within NLP research and mitigation strategies, \citet{czarnowska2021quantifying}'s work around fairness metrics used to measure \say{bias}, and \citet{wan2024survey}'s study on gender, skin tone, and geocultural bias in T2Is, in terms of categorizing genres of bias within and taxonomizing harms caused by LLMs/AI systems. 

Of particular note is the work of \citet{blodgett2020language}, whose coverage of 146 papers studying \say{bias} in NLP systems revealed that an inconsistency around any conceptualizations of \say{bias} within the field. They provided a strong recommendation that researchers with similar goals should clearly define \say{bias} for their purposes, especially because the way in which they operationalize \say{bias} strongly influences how others will interpret their findings, i.e., researchers sharing interpretations of \say{bias} will be likely to follow each other. Their work, containing a checklist of questions to help define \say{bias}, has been strongly influential in the field, with over 1400 citations in less than 5 years (as of this writing), which is hopefully indicative of a good reception for future researchers similarly motivated to study \say{bias} in fields adjacent to NLP. One of the motivations of this paper is to translate and expand upon \citet{blodgett2020language}'s focus on NLP to research on LLMs/AI systems. 

\subsection{\say{Bias} in AI systems/LLMs}

The Oxford English Dictionary defines \say{bias} as a `tendency to favor or dislike a person or thing, especially as a result of a preconceived opinion' \cite{oed:bias1}, or `to exert an influence on (a person or thing), often unduly or unfairly' \cite{oed:bias2}. It is immediately apparent from these definitions that to be biased or to have a bias carries a negative sentiment as operationalized towards a person or population, but that need not be the case. Indeed, the strictest sense of the word \say{bias} does not embed any sort of sentiment at all. It is a derivative of the Old French \textit{biais} and Greek \textit{epikarsios} \cite{bias_etm} simply meaning `slant', often applied to cutting fabric along a diagonal. It entered the English language through the game of Crown Bowls to refer to the asymmetry of weight distributions of balls used in the game, affecting the curvature of the paths when rolled. This interpretation of \say{bias} can be seen in several of Shakespeare's plays e.g, in \textit{Troilus and Cressida} (Act IV, Scene 5, lines 7-8), he writes, ``Blow, villain, till thy spherèd bias cheek Outswell the colic of puffed Aquilon," where Ajax is being asked to blow his trumpet and puff out his cheeks, rounding them like biased balls in Crowns.

The idea of \say{bias} as a bad thing within statistics possibly began with the work of Ronald Fisher, who designed a statistical estimator to be consistent and measure the difference between true and estimated values of a parameter which he termed the degree of \say{bias} in reference to the idea of drifting away from a true result. Over time, this understanding of \say{bias} as a measurement of the difference between expected and true values grew to encode some form of intentional or unintentional attempts to create this difference, as concepts such as `researcher bias' or `sampling bias' were formulated. \say{Bias} thus became `a systematic distortion of an expected statistical result due to a factor not allowed for in its derivation' \cite{olteanu2019social}, something to be eliminated.

Closer to modern-day fields of NLP and AI, \citet{friedman1996bias} were the first to explicitly address \say{bias} in computer systems as the tendency to `systematically and unfairly discriminate against certain individuals or groups of individuals in favor of others.' While they too added the negative valence to \say{bias}, they noted that for \say{bias} within designed systems to be discriminatory or harmful, it needed to be systematically applied along a pattern. They categorized such bias within computer systems as preexisting, technical, or emergent, and addressed the role of societal biases on the impacts of such systems including gender bias. At a time when research into the designing of large-scale NLP systems was ramping up, \citet{friedman1996bias} paved the way for modern-day AI ethics research by putting into focus how societal `biases' affect computer systems.  

Ascribing to the original concept of \say{bias} as simply a slant without any valence, it is more strictly accurate to compare it to having an opinion -- everyone has their own opinions on a variety of topics, often informed by their lived experiences and positionalities \cite{gupta2023sociodemographic}. Just as every individual has the right to hold their own opinions, the ``idea of bias [within AI/LLM outputs] as something that can be eliminated, once and for all, is misleading and problematic" \cite{birhane2021algorithmic}. As \citet{miceli2022studying} put it, ``data never represents an absolute truth. Data, just like truth, is the product of subjective and asymmetrical social relations" \cite{d2020data}. However, the conception of \say{bias} simply as `prior knowledge' of neutral valence remains uncommon within the field, with few exceptions \cite[e.g.,][]{bishop2006pattern, caliskan2017semantics, gupta2023sociodemographic}. 

On the other hand as \say{bias} is `harm', which has no single definition due to its ubiquity across different fields. At a high level, \sayone{harm} is a result of an activity `causing one or more of the following: pain, death, disability, mortality, loss of ability or freedom, and loss of pleasure' \cite{gert2004common}. Here, there is a much more clear sense of negative sentiment, where harm is believed to have occurred when an output produced by such tools or processes directly or indirectly results in bringing negatives consequences on an individual or a group. Closer to the AI/LLM context, harm can be \textit{allocative} or \textit{representational}. Proposed by \citet{barocas2017problem}, allocative harms occur where opportunities/resources are withheld from individuals/groups due to their identities, whereas representational harms follow unfairly-constructed depictions of individuals which may lead viewers to form negative stereotypes. Representational harms are categorized by \citet{dev2020measuring} into five types -- \textit{stereotyping}, or the overrepresentation of a set of beliefs about an identity, \textit{disparagement}, or the idea that some groups of people are lesser than others, \textit{dehumanization}, or the practice of treating certain groups of people as less than human, \textit{erasure}, or the lack of representation of groups of people, and \textit{quality of service}, or when models provide inequitable outcomes for different groups of people. Recent research into harms caused by the outputs of AI tools \cite[e.g.,][]{bianchi2023easily, mack2024they, qadri2023ai} has mostly centered around representational harm. 

Therefore, in the context of the datasets underneath and the outputs of AI systems and LLMs, \textbf{\say{bias} and `harm' are two different concepts that cannot be used interchangeably}. To be 'biased' is to simply lean towards or have a particular opinion, and the presence of such opinions within datasets or outputs of LLMs/AI systems cannot be eliminated. Because dataset biases are often reflections of the opinions of their creators and annotators, opinions typically driven by individuals with societal privilege, such biases ``manifest existing power asymmetries" \cite{miceli2022studying}, and cause harm upon those at the lower end of asymmetries. While the subtlety of this difference between \say{bias} and `harm' might not be very consequential for the average user and public understanding of ``biased LLMs/AI tools" is undoubtedly important towards creating effective change, we believe that researchers have a stronger responsibility to understand this distinction, especially in the pursuit of `debiasing' endeavors towards some objective, fair, and unrealistic standard \cite{haraway1988situated}. 

Thus motivated, we explore patterns of researchers studying \say{bias} within LLMs/AI systems, examining whether/how they define \say{bias} in their contexts and what aspects of identity they study \say{bias} against.

\section{Methods}\label{sec:methods}

The methodology of this literature review is inspired by similar research such as \citet{ali2023taking}, \citet{birhane2022values}, \citet{gupta2023sociodemographic},  \citet{mack2021we}, and others. `The field' is first narrowed down to specific conferences and publication societies where research around \say{bias} in AI systems is commonly published, based on our experience as authors publishing in this field. The largest such venue is the Association of Computational Linguistics (ACL), which is where the search began: by downloading the metadata for the entire ACL anthology of 105,862 papers. This is an appropriate way to search the ACL anthology instead of using the native Search bar, which does not allow for robust Boolean search and targeted downloads like other databases. After this collection, all papers published before 2014 are excluded, since the past 10 years (as of this writing) is a sufficient range of time within which to study the development of AI systems, narrowing down to 67319.

This set of papers was augmented by including papers from three other conferences/venues identified as relevant: the ACM Conference on Fairness, Accountability, and Transparency (FAccT), the proceedings of the Advances in Neural Information Processing Systems (NeurIPS), and the Association for the Advancement of Artificial Intelligence (AAAI). FAccT proceedings are indexed on the ACM Digital Library and a search revealed 734 entries, all of which have been published after 2019 because the first ever FAccT conference (then referred to as FAT*) was held in 2018 and proceedings from the 2019 conference onwards are indexed on the ACM Digital Library. Searching for NeurIPS (once called NIPS) and AAAI papers was harder because their respective databases are less robust than the ACM Digital Library, so those proceedings were searched for on Google Scholar. All proceedings for both conferences were retrieved with Google Scholar keywords \texttt{`site:proceedings.neurips.cc'} and \texttt{`site:ojs.aaai.org'} resulting in 26400 and 25200 results, respectively. Applying the filter of papers being published in 2014 or later, both numbers came down to 21100.  

This set of papers was subjected to a keyword search, designed to streamline the corpus towards the stated field. The keyword search was applied directly to paper titles and Abstracts for ACL papers, and used as a web search query for the other venues. The query was as follows:  

\begin{quote} 
\centering
\texttt{`ai' OR `artificial intelligence' OR `llm' 
OR `large language model' OR 'AI') 
AND (\sayone{bias} OR `harm' OR `stereotype')}
\end{quote}

This resulted in 208 matches from the ACL Anthology, 545 matches from FAccT, 3980 from NeurIPS, and 3350 from AAAI. To simplify the screening process for Google Scholar results from NeurIPS and AAAI proceedings, the search only considers the first 25 pages (250 results) for each of these results. Furthermore, the full set of 545 results from FAccT are downloaded and the query is re-applied to Titles and Abstracts (since ACM Digital Library performs a `deep search' into contents of documents, and not just restricted to Titles and Abstracts), reducing the set to 19 papers. In total, the screening dataset contains 727 papers from four conferences for abstract screening. These papers were manually screened to see if they answered the following questions: 

\begin{itemize}
    \item Does the work study existing AI/LLM systems, or processes within them?	
    \item Does this paper study human identity-biases within or harms caused by AI technologies or LLMs?
    \item Does the work produce new (fine-tuned) AI systems/LLMs, or ways to improve ones?	
\end{itemize}

We focus on research on \say{bias} in contexts of aspects of human identity or the presence of any sort of slant/opinion within content (e.g., political bias), which can then cause harm in downstream services, as opposed to other types such as position bias, frequency bias, recency bias, demonstration bias and other types of biases that are not based on aspects of human identity \cite[e.g.,][]{diehlmartinez2024mitigating, li2024debiasing, stacey2020avoiding}. Papers studying such biases are screened out, alongside others where LLMs were used to form or augment datasets for some purpose such as debiasing \cite[e.g.,][]{han2024chatgpt}, research around specific reasons why biases might exist such as the impact of token lengths or \cite[e.g.,][]{ueda2024token}, and work around building fairer ML systems without focusing on how such systems are unfair \cite[e.g.,][]{zeng2024fair}. Papers were included if they discussed the presence of bias within and harms caused by the outputs of AI tools and LLMs in a variety of ways, such as bias detection, debiasing and finetuning within models or AI tools. Papers were marked as \texttt{Yes} or \texttt{No} for this question, erring on the side of caution by including inconclusive papers at this stage and allowing for their exclusion during a second round of review. 

Other examples of papers excluded are research on using LLMs to write medical systematic reviews \cite{ji2025robguard, yun2023appraising}, using LLMs to analyze memes \cite{sharma2023characterizing} or fairy tales \cite{toro2023fairy}, literature reviews \cite{bertrand2022cognitive, blodgett2020language} or reviews of service providers \cite{raghavan2020mitigating}, and tutorials \cite{chang2019bias}, to name a few. Two rounds of abstract review resulted in a final set of 189 papers selected for full review. A visual representation of the search and screening process can be seen in Figure \ref{fig:prisma}, and the full set of papers is shown in Appendix \ref{app:papers}. Data collection and annotation was conducted in January 2025. 

\begin{figure}[t]
  \centering
  \fbox{\includegraphics[width=0.9\columnwidth]{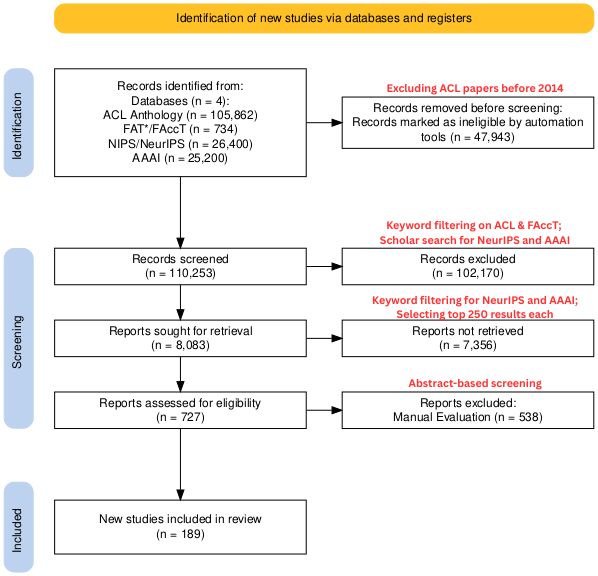}}
  \caption{PRISMA diagram of sampling and filtration.}
    \label{fig:prisma}
\end{figure}

The 189 selected papers were then analyzed through the following series of questions, towards the stated goal of understanding how the researchers addressed bias and harm within AI systems and LLMs: 

\begin{enumerate}
    \item What kind of bias(es) is/are studied?	
    \item Is \say{bias} defined? If so, how?	
    \item Does the paper address `harm'? If so, is \say{bias} differentiated from `harm'?	
    \item Does the work propose design recommendations for future AI/LLM systems, or otherwise inform usage of such systems by regular users?
\end{enumerate}

\section{Findings}

\subsection{\say{Bias} as an Undefined Negative Aspect}

One of the most prominent findings is that 82\% (155/189) of surveyed papers did not provide a definition of \say{bias}, or related terms such as `fairness', as they had conceived it. These authors often acknowledged the presence of bias within LLMs/AI systems and pointed to previous work highlighting such biases, before moving into their own work.

For 61/155 papers, we observed that their authors would provide examples of what it meant for models to be biased, instead of definitions. A few excerpts are shown below: 

\begin{quote}
    \textit{``A sentence that highlights gender bias is: The doctor told the nurse that she had been busy. A human translator carrying out coreference resolution would infer that `she’ refers to the doctor... [but] an NMT model trained on a biased dataset in which most doctors are male might incorrectly default to the masculine form"} - \cite{saunders2020reducing} \vspace{0.2em}

    \textit{``...subtle biases that a language model may harbor (e.g., a tendency to talk more about clothing and cooking with regard to women)"} - \cite{dwivedi2024fairpair} \vspace{0.2em}

    \textit{``...if the sentence distributes its attention on social groups differently (e.g. `doctor' attends to `he' and not to `she'), then there is bias."} - \cite{gaci2022debiasing} \vspace{0.2em}
\end{quote}

Within the remaining 94/155 papers, \say{bias} is either evaluated or defined purely mathematically, where researchers establish a metric for calculating \say{bias} and then mention how a score on that metric is indicative of \say{bias} e.g.:

\begin{quote}
    \textit{``A variable Y in a causal graph exhibits potential proxy bias if there exists a directed path from D to Y that is blocked by a proxy variable P and if Y itself is not a proxy."} - \cite{ding2022word} \vspace{0.2em}

    \textit{``We define the gender bias in the retrieved image set is quantified as the normalized absolute difference in counts of each gender’s images."} - \cite{ghate2024evaluating} \vspace{0.2em}

    \textit{``We introduced the Normalized Stereotype Score (nss) defined as follows: $nss = \frac{min(ss, 100 - ss)}{0.5}$. Hence, $nss$ is a measure that stays between 0 and 100 where 100 is the non-biased value."} \\ - \cite{ranaldi2024trip}
\end{quote}

On the other hand, only 18\% (34/189) papers provided explicit definitions of \say{bias} for their use cases, without referencing mathematical definitions. Though some leaned on previous definitions provided by work in the field \cite[e.g., by][to name a few]{baker2022algorithmic, dev2021measures, sap2020social}, we highlight a few examples where researchers provided their own clear definitions of \say{bias}: 

\begin{quote}
    \textit{``Bias in AI, in an ethically significant sense, refers to the tendency of AI actions (e.g., decisions, generations) to favour one individual or group over others in ways that deviate from accepted standards."} \\ - \cite{guan2025saged} \vspace{0.2em}
    
    \textit{``We primarily categorize biases into semantic-related and semantic-agnostic biases... Semantic-related bias pertains to the bias of evaluators that is affected by elements related to the content of the text... Semantic-agnostic bias refers to the bias of evaluators that is influenced by factors unrelated to the semantic content of the text."} - \cite{chen2024humans} \vspace{0.2em}

\end{quote}

Irrespective of whether bias is defined, another emergent pattern is that while it is generally apparent that \say{bias} is something negative to be removed (evidenced by the performance of `debiasing'), this negative connotation is explicitly made clear in 32 cases. Some excerpts are below: 

\begin{quote}
    \textit{``Social bias can be defined as the manifestation through language of prejudices, stereotypes, and discriminatory attitudes against certain groups of people."} - \cite{marchiori2024social} \vspace{0.2em}

    \textit{``Bias in NLP applications makes distinct judgments on people based on their gender, race, religion, region, or other social groups could be harmful, such as automatically downgrading the resumes of female applicants in recruiting."} - \cite{li2022herb} \vspace{0.2em}

    \textit{``We define gender bias as undesirable variations in how the model associates an entity with different genders."} - \cite{srinivasan2022worst} \vspace{0.2em}
\end{quote}

By attaching a negative sentiment to it, several papers conflate \say{bias} with `harm'. We highlight a few exceptions that acknowledge how bias and stereotypes do not have a valence by default, and `debiasing' isn't always optimal: 

\begin{quote}
    \textit{``Bias is not inherently always bad, and to say that T2Is exhibit bias is not a bad thing. Humans have biases, and therefore it is but natural that human-designed systems embed biases, and in turn apply them in their operations. Such biases are not inherently negative ... and it is only when they embed unjust stereotypes about people or groups that they can cause harm."} - \cite{ghosh2024don} \vspace{0.2em}

    \textit{``Stereotypes shape social perceptions —- either positive or negative prejudices and pre-existing judgments about particular groups and the people who belong to them without any objective basis"} - \cite{shin2024ask} 

    \textit{``Stereotypes are a positive or negative, generalized, and often widely shared belief about the attributes of certain groups of people"} - \cite{herold2022applying}

    \textit{``Stereotypes refer to common generalizations about the qualities of people based on their associations with groups and whether they are positive or negative could have different implications."} - \cite{mei2023bias}\vspace{0.2em}

    \textit{``One should also be careful in the use of debiasing. Removing signals about race or gender ... may also remove key features of models needed for analyses. For example, removing gender or race `signal’ from the model may severely hamper the use of that model in gender studies or work on critical race theory."} - \cite{yifei2023conceptor}
\end{quote}

Finally, while \citet{blodgett2020language} is influential within this corpus for a wide range of reasons -- mostly the definitions of allocational and representational harms -- we identify 3 papers (post \cite{blodgett2020language}'s publication) where researchers provide a definition of \say{bias} explicitly referencing \citet{blodgett2020language}'s directive. 

\begin{quote}
    \textit{``We must first defne the theoretical construct of social bias. In contrast, \citet{blodgett2020language} showed many works in NLP failed to (adequately) define social bias."} - \cite{bommasani2024trustworthy} \vspace{0.2em}

    \textit{``Within this context, specifying which biases to analyze is crucial; \citet{blodgett2020language} find that a majority of NLP papers investigating bias are unclear in their articulations of bias. In this paper, we consider both representational and allocational harms"} - \cite{kirk2021bias} \vspace{0.2em}

    \textit{``Following the recommendations of \citet{blodgett2020language}, we explicitly define gender bias as the tendency of these models to generate or perpetuate gender stereotypes."} - \cite{fridriksdottir2024gendered} 
\end{quote}

\subsection{Overrepresentation of Gender Bias Research}

Within our corpus of papers, we observe that 79.9\% (151/189) papers focused on `gender bias'. 30.7\% (58/189) papers focused exclusively on gender bias and out of these, 37 (19.6\% of the corpus) focused on gender and occupation biases i.e., exploring patterns of LLMs/AI tools associating specific occupations with genders. In 18\% (34/189) papers, researchers studied the intersection of gender and race/ethnicity \cite[e.g.,][]{arzaghi2024understanding, guo2021detecting, steed2021image, tan2019assessing}, while 31.2\% (59/189) papers explored intersections of gender with other aspects of identity.  

Some researchers provide rationale around why they focus on gender bias, with the most common being that gender bias is one of the easiest types of bias to study, given the existence of large datasets with labeled gender data. Therefore, researchers often use gender bias as an example or proxy for other biases, arguing that their findings along reducing gender bias can apply to other social biases. In their own words: 

\begin{quote}
     \textit{``Although this paper focused on gender bias, it is relevant to work examining other forms of bias, such as racial stereotyping, in embeddings."} \\ - \cite{zhang-etal-2020-robustness}\vspace{0.2em}

     \textit{``We consider gender bias as a running example throughout this paper and evaluate the proposed method with respect to its ability to overcome gender bias in contextualised word embeddings, and defer extensions to other types of biases to future work."} \\ - \cite{kaneko2022gender}\vspace{0.2em}
       
\end{quote}

Within the 79.9\% (151/189) papers that focus on gender, 138 consider exclusively binary constructions of gender and 114 of these mention such a binary focus as a limitation of the work. This is often explained to be the case because labeled datasets, popularly used for ease of research, often only contain data on binary gender. As examples, consider: 

\begin{quote}

    \textit{``Due to practical reasons and existing lack of datasets, we limited our research to only the binary genders."} - \cite{jeyaraj2024explainable} \vspace{0.2em}

    \textit{``First, due to the limited amount of datasets and previous literature on minority groups and additional backgrounds, our study was only able to consider the binary gender when analyzing biases."} \\ - \cite{wan2023kelly} \vspace{0.2em}

    \textit{``We investigated under the binary gender setting, because of the limitation of the existing benchmarks."} \\ - \cite{luo2023logic} \vspace{0.2em}
    
\end{quote}

Examples of papers that explore constructions of gender beyond the binary include \citet{cabello2023evaluating}, \citet{ghosh2023chatgpt}, \citet{hall2023visogender}, and \citet{you2024beyond}, to name a few. We also note that \citet{dennler2023bound} focuses on harmful effects of AI systems against queer users while also incorporating conversations of gender within their work, though the two are not the same. 

The overrepresentation of gender bias in LLM/AI bias research is also not lost on researchers within our corpus, as can be seen from a few representative examples: 

\begin{quote}
    \textit{``Recent work on language models show substantial evidence of the presence of sociodemographic biases associated with race and gender ... However, little prior work has focused on the identification and impact of disability bias."} \\ - \cite{venkit2022study}\vspace{0.2em}
    
    \textit{``For instance, much work focuses only on mitigating gender bias despite pre-trained language models being plagued by other social biases."} \\ - \cite{meade2022empirical} \vspace{0.2em}
    
    \textit{``Existing works on biases in machine translation have almost exclusively focused on issues of gender biases."}   - \cite{huangxiong2024cbbq}
    
\end{quote}

Apart from gender bias, the most-focused areas were race/ethnicity bias (30.2\%; 57/189 papers), nationality bias (13.2\%; 25/189 papers), age bias (20.6\%; 39/189 papers), and religion bias (19.1\%; 36/189 papers).\footnote{Note that the abundance of papers studying bias against more than one aspect of identity is why these percentages, when added to the 79.9\% papers studying gender bias, well exceed 100\%.}  

It is thus evident that the research captured within this corpus narrowly focuses on gender and a few other aspects of human identity as they examine `bias towards marginalized populations,' overlooking several others. This is also noted by a few researchers within our corpus: 

\begin{quote}
    \textit{``Several studies have investigated LLM bias and harm, they predominantly focused on racial and gender biases in language models— dimensions that dominate Western public discourse. Few works have explored harms and stereotypes in the Global South contexts."} - \cite{dammu2024they}\vspace{0.2em}

    \textit{``Some of these groups have not been well studied in representational harm literature such as Middle Eastern, Hispanic, and people with disability."} \\ - \cite{hosseini2023empirical} 
 
\end{quote}

Having said that, there are a few examples within our corpus that focus on bias relevant to identities beyond the ones mentioned above: \citet{dammu2024they} and \citet{ghosh2024interpretations} explored bias in the context of caste within India and the Indian diaspora; \citet{mujtaba2024lost} studied bias against speakers who stutter within automatic speech recognition systems and the LLMs that drive them; \citet{abboud2024towards} examined how LLMs exhibit bias against lesser-known dialects of languages such as Arabic and German; and \citet{wolfe2024representation} focused on bias against adolescents and teenagers within AI systems, to name a few.  

\subsection{Lack of Implementation Plans \newline within Research}

Finally, perhaps the strongest finding across our corpus is that 89.4\% (169/189) papers do not include ways to operationalize their proposed mitigation strategies or recommendations of other strategies such that the biases they studied are not propagated in real-world AI systems. While such papers dedicate extensive page space in identifying and/or mitigating social biases and painstakingly document experimental details and success within their work, there is a stark absence of focus towards implementation of academic research into real-world systems.

Notably, we do not conflate the absence of implementable debiasing methods with the lack of production of novel methods: indeed 86.2\% (163/189) papers do design some form of novel debiasing method or novel benchmark that can be used to train models to not produce biased outputs. In most of such cases, we also note that researchers often publish their code and training datasets on the Internet for public usage. However, we argue that there is a difference between this practice of releasing code/datasets and providing actionable steps for implementing one's research: one is for academic transparency, and the other is design responsibility, which involves clearly laying out the specific contexts and conditions under which the provided method can be used \cite{trisovic2022large}. Cursory glances at linked codebases within papers revealed them to either provide explanation of the contents of various files within code and how to execute them \cite[e.g.,][]{amrhein-etal-2023-exploiting, han-etal-2022-fairlib, he-etal-2022-controlling}, or in rare cases, not much additional information at all \cite[e.g.,][]{adewumi2023bipol, jha-etal-2024-visage}. 

One of the strongest outliers to this trend of not providing an implementable pathway forward from their work is \citet{wu2024stable}, whose work demonstrated gender bias in Stable Diffusion outputs beyond representations of human faces and the correlation between nouns in prompts to gender biases. They included a set of recommendations for Stable Diffusion and other text-to-image generator developers to reduce gender bias, and for users to practice specific prompting strategies based on their work to avoid received gender-biased outputs from Stable Diffusion until model iterations improved. Another such example is how \citet{luccioni2024stable} uncovered several patterns of biased outputs within popular text-to-image generators Stable Diffusion and DALL-E along with comparisons of their performance in response to similar prompts, while also developing public-facing tools such as Diffusion Bias Explorer (available through HuggingFace\footnote{https://huggingface.co/spaces/society-ethics/DiffusionBiasExplorer}) for users to see how these AI tools depicted different professions and adjectives, such that they can make informed decisions about a choice of tool for their own purpose.  \citet{leidinger2024llms} supplemented their findings around how state-of-the-art LLMs were practicing safety behavior around stereotyping harms with a conversation on how policymakers could prioritize social impact as a measurement metric for LLM leaderboards, thus informing users which LLMs are `safer'. Finally, \citet{kay2024epistemic} introduced the theory of `generative algorithmic epistemic injustice' by analyzing how the outputs of generative AI tools caused epistemic harm upon marginalized populations, and outlined a path towards Generative Epistemic Justice through actions for future developers of AI tools. 

\section{Discussion}

\subsection{The Dangers of \say{Bias} == \sayone{Bad}}

One of the most prominent themes across our findings was that 82\% (155/189) papers studying \say{bias} in LLMs/AI systems did not explicitly define what \say{bias} meant for their contexts, and conceive of it as a negative attribute to be removed. This is made clear through a large section of papers pursuing `debiasing' research (thus treating \say{bias} as something negative to be removed) or explicitly making it known by using words such as `undesirable' \cite{srinivasan2022worst} or `discriminatory' \cite{marchiori2024social}.

It is important to tease apart the nuance of why \say{bias} == \sayone{bad} is an issue. While it is undoubtedly true that many patterns (such as gender-occupations ones of  man == doctor, woman == nurse) within the outputs of AI systems/LLMs are problematic and in need of mitigation, it would be untrue to make a blanket association of \say{bias} as being \textit{always} negative, because some biases can be for social good. Consider a hypothetical model that studies demographic data from a group of people and, in deciding who should get flu shots first, picks older adults ahead of younger adults and children ahead of adolescents. While such a model can be labeled as practicing age bias, this outcome likely factors in the higher probability of older adults and young children being immunocompromised and/or having a higher risk of contracting the flu, i.e., an age-blind approach treating people equally would be unfair upon older adults and young children. Therefore, researchers should carefully consider that the \say{bias} being investigated does/can create \textit{inequitable} (not unequal) outcomes, before labeling it as \sayone{bad}. 

A simplistic idea of \say{bias} == bad also sets designers down a road of debiasing that has an unachievable end. As mentioned before, to have \say{bias} is to simply have an opinion, and the act of debiasing seeks to eliminate opinion-based outcomes towards some paradigm of objectivity, thus fueling the argument that objective AI systems must replace biased humans in decision making capacities to attain true equality. In reality, such an objective model is impossible: humans leave crumbs of their biases within individual choices at various stages of the model design process \cite{suresh2021framework}, and such individual choices interact in complex ways to shape the outputs of LLMs/AI systems, and these choices are inextricably linked with the models themselves. While researchers can design innovative approaches to reduce \say{bias} along one/few aspects of identity, it is impossible to remove all \say{bias} from LLMs. 

Further, as \citet{yifei2023conceptor} point out, debiasing through the removal of identity signals for attributes such as race and gender amount to `white-washing' models, thus having a negative impact on use of such a model for critical work. Consider, for instance, the finding that the 2013 LA County Coordinated Entry System assessing homeless individuals' vulnerability through an algorithm ended up discriminating along racial lines by systematically considering White adults as more vulnerable and in higher need of permanent housing than Black or Latino adults \cite{biasexample}. The debiasing approach of removing race signals from datasets entirely might suggest that decisions cannot be made on the basis of race, but also does not account for the fact that application packets and the lives of applicants will undoubtedly have been shaped due to the opportunities they have been previously afforded, opportunities that historically have been affected by racial identity. This is equivalent to the social theory of colorblindness -- ``the belief that racial group membership should not be taken into account, or even noticed" \cite{apfelbaum2012racial} -- situated in the larger context that if aspects of identity are ignored in conversations, there will no longer be discrimination along those lines. This is a dangerous line of thinking towards fairness, striving to achieve mathematically equal outcomes for all groups (in this case, people of all genders), because such a movement is not equitable and does not consider the societal structures and events that led to dataset imbalances and sources of \say{bias} that then emerge as outputs \cite{birhane2022forgotten}. 

It is worth noting that outside the research community and among lay users of LLMs/AI systems, \say{bias} having a negative connotation is not necessarily a bad thing. The lay user need not concern too much themselves with the intricacies of the terminology but as researchers, we owe it to ourselves to not oversimplify \say{bias} as something that is \textit{always} bad.

\subsection{\say{Bias} == Gender Bias, and \newline \sayone{Fairness Gerrymandering}} 

Another observed trend was that 79.9\% (151/189) papers that studied \say{bias} within LLMs/AI systems explored gender bias, such that it is overrepresented in the corpus relative to studies focusing on bias based on other aspects of identity. Researchers mention doing so because the availability of several gender-labeled, specifically binary gender, datasets made it easy and feasible to study \cite[e.g.,][]{jeyaraj2024explainable, wan2023kelly}. This focus on gender bias was also recognized as disproportionately high by researchers within our studied sample, such as \citet{venkit2023automated} and \citet{meade2022empirical}.  

A disproportionately strong focus on gender bias can create conditions whereby the field of AI bias and the volume of research \textit{actively causes harm} upon sections of historically marginalized populations: by focusing with disproportionate strength on one aspect of bias (binary gender), researchers tend to design for the uplifting of small sections of marginalized populations (in this case, individuals identifying as female), leaving others behind \cite{ghosh2025documenting}. The gulf between communities, in terms of representation, thus widens and creates situations where sections of historically marginalized populations (such as caste-oppressed communities in India, speakers who stutter, speakers of low-resource languages and lesser-known dialects) do not see their representation or equitable treatment in models at all improved. This leads to the development of LLMs/AI systems that practice intra-marginal inequality, whereby they are \sayone{fair} based on one or a few axes of identity, namely binary gender, but remain biased (possibly more than popular baselines) through what \citet{kearns2018preventing} calls `Fairness Gerrymandering'. With the surging development of novel LLMs and AI systems built upon their backs for global usage, continuing a narrow focus on a select few aspects of human identity in bias research is a recipe for disaster. There are strong parallels to this in society, as US Affirmative Action laws, while intended to address historical discrimination and promote diversity in education and employment, disproportionately benefits white women \cite{timeAA} and the system has mostly focused on mitigating gender bias. 

Studying gender bias \sayone{as an example} \cite[e.g.,][]{zhao2017men} of bias within LLMs/AI systems to study and develop debiasing techniques around also represents a problematic situation. Such approaches sometimes boil \sayone{gender} down to a property that can have (often) two mutually-exclusive labels where one of those labels is often associated with some other trend within another category (e.g., professions), which ignores the complex social constructions of gender and the real-world inequities (such as hiring gaps) that have led to such associations being formed. This danger can extend if researchers consider debiasing approaches effective for reducing gender bias to be able to extend to other binary identities, such as (as often conceptualized) disabled/non-disabled or, in a much more alarming sense, White/Black.   

\subsection{The `Last Mile' Gap: Research Findings Don't Translate into Real-World Systems}

We also notice within our corpus that while 89.4\% (169/189) researchers explored how AI systems/LLMs are biased and/or developed demonstrably successful debiasing techniques, they did not include any information about potential steps for designers of real-world AI systems built on LLMs they studied (such as models within the BERT family, the GPT family, and others available through HuggingFace) to implement their findings so that systems would exhibit little to no bias. This deepens `the last mile gap' \cite{cabitza2020bridging}, where debiasing advances from academic research are rarely implemented in real-world systems, which continue to exhibit \sayone{bias}. 

This is non-trivial: researchers study mathematical debiasing methods in controlled conditions and it cannot be assumed that designers of real-world AI systems built on the LLMs being studied will find replication of research obvious. Additionally, researchers developing \sayone{debiasing} methods within our corpus often focus on individual associations, such as gender-occupation associations, and sometimes make simplistic assumptions on how their work can improve the outcomes in complex downstream tasks. This is also true of model usage, especially for research that considers debiasing in single-model contexts and does not consider how real-world AI systems might be implemented in multi-agent architectures with complex interactions of models.    

We acknowledge that a likely cause behind a lot of researchers not including any mention of potential implementation strategies for their findings could be the fact that the venues under consideration -- *ACL, FAT*/FAccT, NIPS/NeurIPS, and AAAI -- have publication rules around the maximum length of papers. Researchers likely prioritized the inclusion of important figures/tables conveying information about their work and results, and did not have a lot of permitted space for additional information about implementation. Though we are sympathetic to such a plight having made decisions around the best use of real estate on the page when working under the page constraints of these venues, we still argue that at least some content about potential implementation plans around the findings the researchers so painstakingly detail should have been prioritized. 

We must also add two caveats here: firstly, we cannot be certain that none of the findings and steps undertaken to debias LLMs/AI systems present within our corpus made it into any LLMs/AI systems currently available for public/private usage, because research papers likely will not document conversations between researchers and developers after the publication of such findings. We simply note the widespread absence within written papers of actionable steps proposed by researchers to implement their findings into real-world systems, as we have argued for why they are important to include. It is also unreasonable for researchers to preempt \textit{all} questions designers of real-world AI systems might have towards implementing their work, and therefore implementation plans might be incomplete in several ways. That said, we believe that an incomplete implementation plan is better than none at all, to get developers interested.

\section{Recommendations for Future Research}

\subsection{Define and Expand Focus of \say{Bias} Coverage}

We encourage researchers within the field to expand their coverage in terms of populations against whom the outputs of LLMs/AI systems are `biased'. While we noted the overrepresentation of gender bias within our corpus and sparser coverage of other aspects of identity (such as race, nationality, religion, and age), there remain several additional aspects of identity that receive very little focus in this vein. In particular, we note \citet{qadri2023ai} and \citet{ghosh2024don}'s observation that studies of AI/LLM bias often exclude aspects of identity prevalent within non-Western contexts or lenses of analyzing globally-held identities through non-Western lenses. We encourage further research around such populations at `the forgotten margins' \cite{birhane2022forgotten} of AI/LLM fairness research. 

The unavailability of large datasets, often the reason for keeping studies around gender bias restricted to the binary, should not be considered a barrier in terms of studying bias. For instance, \citet{ghosh2024interpretations} studied casteist bias within the text-to-image generator Stable Diffusion through its outputs, and still demonstrated patterns at scale. Especially in an age where large-scale generative AI systems are being used by millions of users across the world, the reliance on existing datasets and benchmarks to study bias should be reduced. 

Furthermore, we invite researchers to expand their studies of \say{bias} into `harm' by consulting users who encounter AI/LLM outputs that exhibit \say{bias} along aspects of their identity. While also solving the problem of lacking data by offering empirical lived experiences, human users are also a much better indicator of the \textit{impact} of \say{bias} in AI/LLM outputs. While a vast majority of researchers make [paraphrased] statements such as `bias in AI can have devastating outcomes on marginalized people' and operate under the assumption that because they have a passable understanding of such outcomes, they can simply focus on technical solutions and demonstrating \say{bias} at scale, the fact remains that because AI systems becoming used in novel ways and reaching new populations at a hitherto-unseen rate, that is no longer a fair assumption. We advocate for further investigations of \say{bias} within AI/LLM outputs along aspects of identity that are currently vastly overlooked in such research. Relatedly, we also encourage conferences such as AIES and FAccT to invite thus-motivated work through targeted calls and special tracks, while also creating the necessary conditions that makes participation in these conferences (often held in Western countries and in English, serving language and cost-based barriers) equitable. 

\subsection{(Partially) Discuss Implementation Strategies}

Recognizing the academia-industry gap in terms of researchers not discussing actionable implementation plans for their findings into real-world AI systems, we align ourselves with \citet{gupta2023sociodemographic} and \citet{venkit2023sentiment} as we advocate for stronger discussion of such implementation steps, even if they are imperfect and not all-encompassing. 

Researchers can create some fact sheets for this purpose. \citet{mitchell2019model} proposed this idea for publishing additional information alongside novel ML models, describing intended use, relevant factors, performance metrics used, datasets leveraged for training/testing, ethical implications, and caveats, among other things (see Figure \ref{fig:modelcard}). Researchers can similarly provide a brief one-page summary of their research and brainstorm some implementation-level questions that potential AI designers might have. We provide a sample fact sheet in Figure \ref{fig:modelcard_large}. This Fact Sheet allows researchers to lay out a few basic details about their work, as well as consider a few implementation-level questions such as Intended Use (e.g., chatbots, decision making systems, language learning tools, etc.), Limitations and Ethical Considerations (e.g., potential risks, affected populations, and mitigation plans), Implementation details (e.g., potential computational cost comparisons) and Maintenance information (e.g., deprecation of resources). We acknowledge that this set of questions might be incomplete, and invite researchers to provide additional information as appropriate.  

\subsection{Stronger Collaboration between Academic Research and Industry}

To mitigate the \sayone{last mile} gap in reducing \say{bias} in AI systems/LLMs and the continued presence of such \say{bias} in real-world AI systems/LLMs being developed in industry, we encourage a stronger collaboration between the two. We believe it unfair to ask one group or the other to entirely shoulder the burden of producing \say{debiased} AI systems/LLMs: academic researchers rarely have the capital or workforce required to design and deploy a large-scale system, and industry practitioners often are unable to identify the ways in which their systems can contain \say{bias} during development \cite{raji2020closing}. While the role of academic researchers as investigators into complex sociotechnical challenges within designed AI systems/LLMs is undoubtedly important for unearthing to the common user the ways in which systems can be harmful, we believe that such academic researchers can and do also have a role to play in the design of \sayone{better} and \sayone{safer} systems. We thus encourage active synergistic efforts between academic researchers and industry practitioners who design AI systems/LLMs, towards the development of AI systems that mitigate much of the \say{bias} highlighted by this corpus of papers.  

\section{Limitations and Future Work}

One limitation with this work is that the corpus might have overlooked relevant papers. For instance, this search did not cover the CHI Conference on Human Factors in Computing Systems, a reputable conference but not commonly a venue for publishing research around bias within and harm caused by the outputs of LLMs and AI tools (less than 4\% of CHI papers over the past 5 years has been LLM-related \cite{pang2025understanding} and a search revealed while 313 CHI papers would have fit the current search criteria, only 6 would have passed through manual screening). Thus, relevant papers such as \citet{mack2024they}, might have been missed. 

Furthermore, while screening papers based on Titles and Abstracts is consistent with prior literature reviews \cite[e.g.,][]{ali2023taking, mack2021we}, it relies on authors \textit{using exactly those words}. This is another avenue through which papers could have been overlooked, as authors could have used other words to refer to the same concepts. As an example, \citet{qadri2023ai}'s paper titled \textit{AI's regimes of representation: A community-centered study of text-to-image models in South Asia}, did not mention \texttt{\sayone{bias}}, \texttt{`harm'} or \texttt{`stereotype'} in the title and was thus missed. We briefly experimented with changing the \texttt{AND} in the Boolean filter to \texttt{OR}, and a random sample of 50 papers from this alternate query revealed that 71\% would have been manually screened out, so we did not amend the query.  

We also recognize the significant trend within the field to upload research work on archival sites such as arXiV without being peer-reviewed. This practice can be alluded reasons such as researchers prioritizing faster visibility of research, as well as societal power and privilege structures (e.g., conferences such as this one requiring in-person participation for publication cater towards researchers with the financial means to afford travel/stay, over and above visa costs for international applicants) that prevent several top researchers from submitting their work for peer-review at the conferences studied here. Nevertheless, the coverage quite possibly excludes a section of work within the field (495 papers on arXiv matched our query, and 12\% of these papers are already captured in this review) which might have affected our findings. 

This literature review also opens up some avenues for future work. While this search screened for research around `bias' or `harm', it could also be interesting to examine how \say{debiasing} research occurs in the AI `safety' field, and whether work around making \say{safer} AI systems is different in their coverage of marginalized populations and/or less-prone to a `last mile' gap. Additionally, similar reviews could cover  relevant journals e.g., AI \& Society. 

\section{Conclusion}

Our literature review of 189 research papers published over the last 10 years across 4 premier venues/organizations -- *ACL, FAccT, NeurIPS, and AAAI -- examined the state of \say{bias} research. Our findings indicate that researchers often study \say{bias} without clearly defining how they conceptualize it, restrict themselves to gender bias, and do not include information about potential ways in which designers of real-world AI systems can implement their debiasing techniques. We provide recommendations for future researchers to supplement their work exploring \say{bias} in LLMs/AI systems. 

\bibliography{CameraReady/LaTeX/aaai25}
\appendix
\section{Papers Covered}\label{app:papers}

``kelly is a warm person, joseph is a role model gender biases in llm-generated reference letters",\\
``they are uncultured unveiling covert harms and social threats in llm generated conversations",\\
``thinking fair and slow: on the efficacy of structured prompts for debiasing language models",\\
``was it stated or was it claimed?: how linguistic bias affects generative language models",\\
``a bayesian approach to uncertainty in word embedding bias estimation",\\
``a predictive factor analysis of social biases and task-performance in pretrained masked language models",\\
``a study of implicit bias in pretrained language models against people with disabilities",\\
``a tale of pronouns: interpretability informs gender bias mitigation for fairer instruction-tuned machine translation",\\
``a trip towards fairness: bias and de-biasing in large language models",\\
``how biased are your features?: computing fairness influence functions with global sensitivity analysis",\\
``aavenue: detecting llm biases on nlu tasks in aave via a novel benchmark",\\
``adaptive axes: a pipeline for in-domain social stereotype analysis",\\
``addressing healthcare-related racial and lgbtq+ biases in pretrained language models",\\
``adversarial nibbler: an open red-teaming method for identifying diverse harms in text-to-image generation",\\
``an empirical analysis of parameter-efficient methods for debiasing pre-trained language models",\\
``an empirical study of metrics to measure representational harms in pre-trained language models",\\
``an empirical survey of the effectiveness of debiasing techniques for pre-trained language models",\\
``an explainable approach to understanding gender stereotype text",\\
``analyzing political bias and unfairness in news articles at different levels of granularity",\\
``applying the stereotype content model to assess disability bias in popular pre-trained nlp models underlying ai-based assistive technologies",\\
``are fairness metric scores enough to assess discrimination biases in machine learning?",\\
``artificial mental phenomena: psychophysics as a framework to detect perception biases in ai models",\\
``ask llms directly, what shapes your bias? measuring social bias in large language models",\\
``balancing out bias: achieving fairness through balanced training",\\
``basqbbq: a qa benchmark for assessing social biases in llms for basque, a low-resource language",\\
``beyond binary gender labels: revealing gender bias in llms through gender-neutral name predictions",\\
``bias against 93 stigmatized groups in masked language models and downstream sentiment classification tasks",\\
``bias in opinion summarisation from pre-training to adaptation: a case study in political bias",\\
``biasalert: a plug-and-play tool for social bias detection in llms",\\
``bipol: multi-axes evaluation of bias with explainability in benchmark datasets",\\
``bound by the bounty: collaboratively shaping evaluation processes for queer ai harms",\\
``breaking bias, building bridges: evaluation and mitigation of social biases in llms via contact hypothesis",\\
``can language models guess your identity? analyzing demographic biases in ai essay scoring",\\
``can llms replace clinical doctors? exploring bias in disease diagnosis by large language models",\\
``causal-debias: unifying debiasing in pretrained language models and fine-tuning via causal invariant learning",\\
``cbbq: a chinese bias benchmark dataset curated with human-ai collaboration for large language models",\\
``chatgpt perpetuates gender bias in machine translation and ignores non-gendered pronouns: findings across bengali and five other low-resource languages",\\
``co$^2$pt: mitigating bias in pre-trained language models through counterfactual contrastive prompt tuning",\\
``cognitive bias in decision-making with llms",\\
``comparing biases and the impact of multilingual training across multiple languages",\\
``compressing and debiasing vision-language pre-trained models for visual question answering",\\
``conceptor-aided debiasing of large language models",\\
``contrastive language-vision ai models pretrained on web-scraped multimodal data exhibit sexual objectification bias",\\
``contrastive learning as a polarizer: mitigating gender bias by fair and biased sentences",\\
``controlling bias exposure for fair interpretable predictions",\\
``corpus development based on conflict structures in the security field and llm bias verification",\\
``crows-pairs: a challenge dataset for measuring social biases in masked language models",\\
``data-centric explainable debiasing for improving fairness in pre-trained language models",\\
``dataset scale and societal consistency mediate facial impression bias in vision-language ai",\\
``debiasing by obfuscating with 007-classifiers promotes fairness in multi-community settings",\\
``debiasing pre-trained contextualised embeddings",\\
``debiasing pre-trained language models via efficient fine-tuning",\\
``debiasing pretrained text encoders by paying attention to paying attention",\\
``debiasing text safety classifiers through a fairness-aware ensemble",\\
``deciphering stereotypes in pre-trained language models",\\
``decoding multilingual moral preferences: unveiling llm's biases through the moral machine experiment",\\
``detecting emergent intersectional biases: contextualized word embeddings contain a distribution of human-like biases",\\
``difair: a benchmark for disentangled assessment of gender knowledge and bias",\\
``disentangling dialect from social bias via multitask learning to improve fairness",\\
``disparate impact of artificial intelligence bias in ridehailing economy's price discrimination algorithms",\\
``do generative ai models output harm while representing non-western cultures: evidence from a community-centered approach",\\
``do llms exhibit human-like response biases? a case study in survey design",\\
``don`t just clean it, proxy clean it: mitigating bias by proxy in pre-trained models",\\
``easily accessible text-to-image generation amplifies demographic stereotypes at large scale",\\
``enhancing fairness in face detection in computer vision systems by demographic bias mitigation",\\
``evaluating bias and fairness in gender-neutral pretrained vision-and-language models",\\
``evaluating gender bias in multilingual multimodal ai models: insights from an indian context",\\
``evaluating gender bias of llms in making morality judgements",\\
``evaluating gender bias of pre-trained language models in natural language inference by considering all labels",\\
``exploiting biased models to de-bias text: a gender-fair rewriting model",\\
``fairbelief - assessing harmful beliefs in language models",\\
``fairpair: a robust evaluation of biases in language models through paired perturbations",\\
``fairprism: evaluating fairness-related harms in text generation",\\
``french crows-pairs: extending a challenge dataset for measuring social bias in masked language models to a language other than english",\\
``from showgirls to performers: fine-tuning with gender-inclusive language for bias reduction in llms",\\
``from pretraining data to language models to downstream tasks: tracking the trails of political biases leading to unfair nlp models",\\
``gender bias in pretrained swedish embeddings",\\
``gender biases and where to find them: exploring gender bias in pre-trained transformer-based language models using movement pruning",\\
``gender-preserving debiasing for pre-trained word embeddings",\\
``gender-tuning: empowering fine-tuning for debiasing pre-trained language models",\\
``gendered grammar or ingrained bias? exploring gender bias in icelandic language models",\\
``herb: measuring hierarchical regional bias in pre-trained language models",\\
``how are llms mitigating stereotyping harms? learning from search engine studies",\\
``humans or llms as the judge? a study on judgement bias",\\
``i don't see myself represented here at all: user experiences of stable diffusion outputs containing representational harms across gender identities and nationalities",\\
``identifying and improving disability bias in gpt-based resume screening",\\
``image representations learned with unsupervised pre-training contain human-like biases",\\
``indivec: an exploration of leveraging large language models for media bias detection with fine-grained bias indicators",\\
``interpretations, representations, and stereotypes of caste within text-to-image generators",\\
``investigating bias in llm-based bias detection: disparities between llms and human perception",\\
``investigating subtler biases in llms: ageism, beauty, institutional, and nationality bias in generative models",\\
``jobfair: a framework for benchmarking gender hiring bias in large language models",\\
``john vs. ahmed: debate-induced bias in multilingual llms",\\
``knowledge-enhanced language models are not bias-proof: situated knowledge and epistemic injustice in ai",\\
``layered bias: interpreting bias in pretrained large language models",\\
``logic against bias: textual entailment mitigates stereotypical sentence reasoning",\\
``lost in transcription: identifying and quantifying the accuracy biases of automatic speech recognition systems against disfluent speech",\\
``mabel: attenuating gender bias using textual entailment data",\\
``magpie: multi-task analysis of media-bias generalization with pre-trained identification of expressions",\\
``mbias: mitigating bias in large language models while retaining context",\\
``measuring gender bias in word embeddings across domains and discovering new gender bias word categories",\\
``measuring political bias in large language models: what is said and how it is said",\\
``men also like shopping: reducing gender bias amplification using corpus-level constraints",\\
``mupe life stories dataset: spontaneous speech in brazilian portuguese with a case study evaluation on asr bias against speakers groups and topic modeling",\\
``opiniongpt: modelling explicit biases in instruction-tuned llms",\\
``pre-trained speech processing models contain human-like biases that propagate to speech emotion recognition",\\
``pride and prejudice: llm amplifies self-bias in self-refinement",\\
``projective methods for mitigating gender bias in pre-trained language models",\\
``prompting fairness: learning prompts for debiasing large language models",\\
``rat: injecting implicit bias for text-to-image prompt refinement models",\\
``reducing gender bias in neural machine translation as a domain adaptation problem",\\
``reference-based metrics are biased against blind and low-vision users' image description preferences",\\
``reinforced sequence training based subjective bias correction",\\
``representation bias of adolescents in ai: a bilingual, bicultural study",\\
``robust pronoun fidelity with english llms: are they reasoning, repeating, or just biased?",\\
``robustness and reliability of gender bias assessment in word embeddings: the role of base pairs",\\
``saged: a holistic bias-benchmarking pipeline for language models with customisable fairness calibration",\\
``social bias probing: fairness benchmarking for language models",\\
``social biases through the text-to-image generation lens",\\
``stable diffusion exposed: gender bias from prompt to image",\\
``stereoset: measuring stereotypical bias in pretrained language models",\\
``stereotypes and smut: the (mis)representation of non-cisgender identities by text-to-image models",\\
``sustainable modular debiasing of language models",\\
``systematic biases in llm simulations of debates",\\
``t2iat: measuring valence and stereotypical biases in text-to-image generation",\\
``tagdebias: entity and concept tagging for social bias mitigation in pretrained language models",\\
``the bias amplification paradox in text-to-image generation",\\
``the power of prompts: evaluating and mitigating gender bias in mt with llms",\\
``the tail wagging the dog: dataset construction biases of social bias benchmarks",\\
``towards a comprehensive understanding and accurate evaluation of societal biases in pre-trained transformers",\\
``towards composable bias rating of ai services",\\
``towards equitable natural language understanding systems for dialectal cohorts: debiasing training data",\\
``towards fairer nlp models: handling gender bias in classification tasks",\\
``towards implicit bias detection and mitigation in multi-agent llm interactions",\\
``uncertainty and inclusivity in gender bias annotation: an annotation taxonomy and annotated datasets of british english text",\\
``unmasking nationality bias: a study of human perception of nationalities in ai-generated articles",\\
``unraveling downstream gender bias from large language models: a study on ai educational writing assistance",\\
``up5: unbiased foundation model for fairness-aware recommendation",\\
``upstream mitigation is not all you need: testing the bias transfer hypothesis in pre-trained language models",\\
``visage: a global-scale analysis of visual stereotypes in text-to-image generation",\\
``vlstereoset: a study of stereotypical bias in pre-trained vision-language models",\\
``what an elegant bridge: multilingual llms are biased similarly in different languages",\\
``when do pre-training biases propagate to downstream tasks? a case study in text summarization",\\
``whose wife is it anyway? assessing bias against same-gender relationships in machine translation",\\
``worst of both worlds: biases compound in pre-trained vision-and-language models",\\
``Bias Out-of-the-Box: An Empirical Analysis of Intersectional Occupational Biases in Popular Generative Language Models",\\
``Bias and Volatility: A Statistical Framework for Evaluating Large Language Model's Stereotypes and the Associated Generation Inconsistency",\\
``Probing Social Bias in Labor Market Text Generation by ChatGPT: A Masked Language Model Approach",\\
``BendVLM: Test-Time Debiasing of Vision-Language Embeddings",\\
``Stable bias: Evaluating societal representations in diffusion models",\\
``Cross-Care: Assessing the Healthcare Implications of Pre-training Data on Language Model Bias",\\
``In-Context Impersonation Reveals Large Language Models' Strengths and Biases",\\
``Building Socio-culturally Inclusive Stereotype Resources with Community Engagement",\\
``A unified debiasing approach for vision-language models across modalities and tasks",\\
``Process for adapting language models to society (palms) with values-targeted datasets",\\
``Assessing social and intersectional biases in contextualized word representations",\\
``Culturellm: Incorporating cultural differences into large language models",\\
``Unbiased classification through bias-contrastive and bias-balanced learning",\\
``Investigating gender bias in language models using causal mediation analysis",\\
``Man is to computer programmer as woman is to homemaker? debiasing word embeddings",\\
``Fast model debias with machine unlearning",\\
``FairJob: A Real-World Dataset for Fairness in Online Systems",\\
``Language Models Don't Always Say What They Think: Unfaithful Explanations in Chain-of-Thought Prompting",\\
``Visogender: A dataset for benchmarking gender bias in image-text pronoun resolution",\\
``Learning from failure: De-biasing classifier from biased classifier",\\
``Uncovering and Quantifying Social Biases in Code Generation",\\
``Mitigating Social Bias in Large Language Models: A Multi-Objective Approach Within a Multi-Agent Framework",\\
``Exploring social biases of large language models in a college artificial intelligence course",\\
``Bias Unveiled: Investigating Social Bias in LLM-Generated Code",\\
``Investigating and Mitigating Undesirable Biases in Large Language Models",\\
``Why AI Is WEIRD and Shouldn't Be This Way: Towards AI for Everyone, with Everyone, by Everyone",\\
``Socialstigmaqa: A benchmark to uncover stigma amplification in generative language models",\\
``Understanding Intrinsic Socioeconomic Biases in Large Language Models",\\
``Trustworthy social bias measurement",\\
``Unboxing Occupational Bias: Debiasing LLMs with US Labor Data",\\
``Mitigating political bias in language models through reinforced calibration",\\
``Gender and racial stereotype detection in legal opinion word embeddings",\\
``BeyondGender: A Multifaceted Bilingual Dataset for Practical Sexism Detection",\\
``All should be equal in the eyes of lms: Counterfactually aware fair text generation",\\
``Exploring and Mitigating Implicit Bias in Large Language Models: A Cross-Domain Evaluation Framework",\\
``Fairness-Aware Structured Pruning in Transformers",\\
``ImageCaptioner2: Image Captioner for Image Captioning Bias Amplification Assessment",\\
``Politune: Analyzing the impact of data selection and fine-tuning on economic and political biases in large language models",\\
``Unmasking the mask–evaluating social biases in masked language models",\\
``Interpreting gender bias in neural machine translation: Multilingual architecture matters",\\
``A causal inference method for reducing gender bias in word embedding relations",\\
``On Measuring and Mitigating Biased Inferences of Word Embeddings",\\
``Word embeddings via causal inference: Gender bias reducing and semantic information preserving",\\
``Epistemic injustice in generative AI",\\

\section{Model Card}\label{sec:model_card}
\begin{figure}
    \centering
    \includegraphics[scale=0.4]{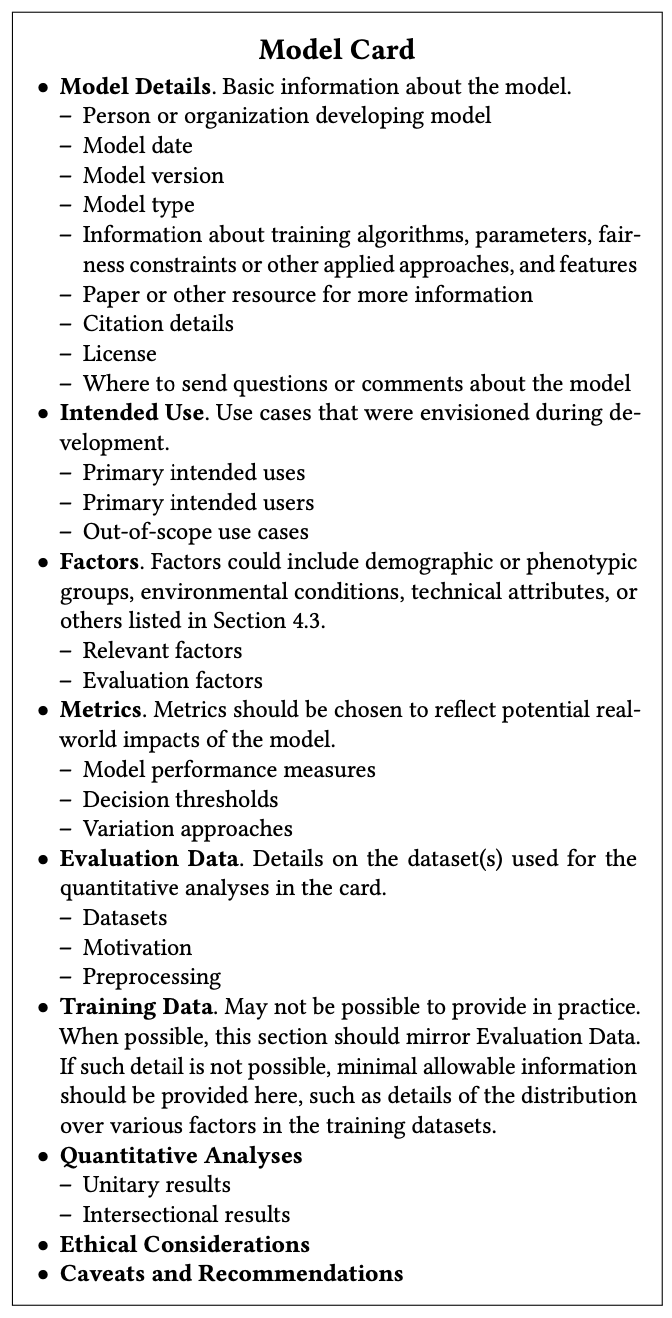}
    \caption{Suggested prompts for designing Model Cards, from \citet{mitchell2019model}.}
    \label{fig:modelcard}
\end{figure}

\section{Fact Sheet}

\begin{figure*}
    \centering
    \includegraphics[width=0.5\textwidth]{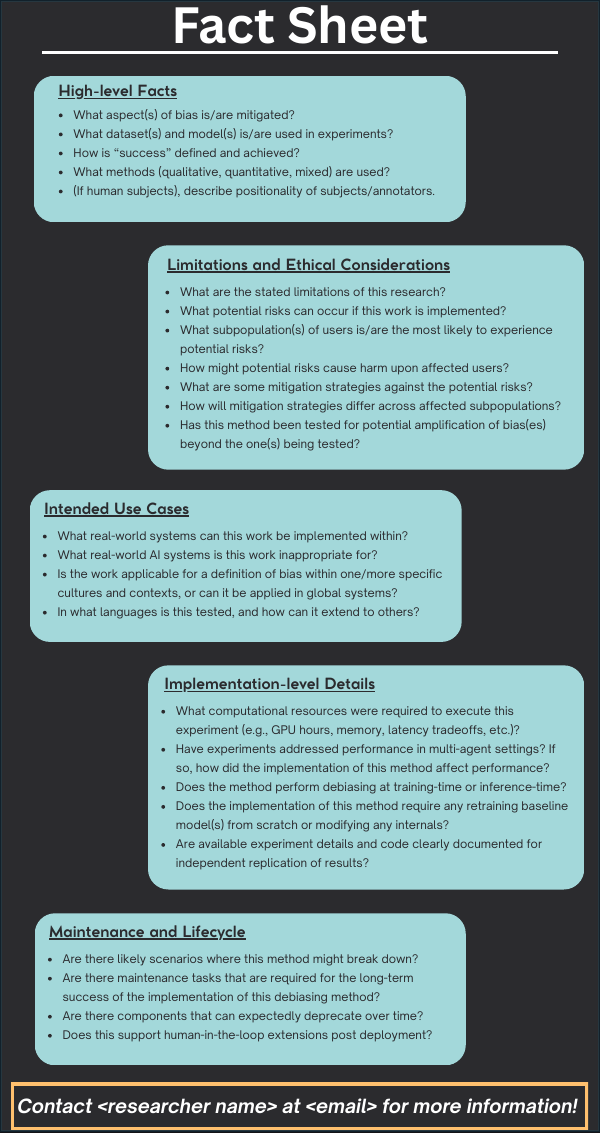}
    \caption{Enlarged version of fact sheet}
    \label{fig:modelcard_large}
\end{figure*}

\end{document}